 \documentclass[final,3p,times,number]{elsarticle}

\usepackage{amssymb}
\usepackage{soul}
\usepackage{lipsum}
\usepackage{xcolor}
\usepackage{mathtools, cuted}
\usepackage[english]{babel}
\usepackage{amsmath}
\usepackage{hyperref}
\usepackage{amsthm}
\usepackage{tikz}
\journal{Physics Letters B}

\begin{document}

\begin{frontmatter}



\title{Testing the cosmological Poisson equation in a model-independent way}


\author[first]{Ziyang Zheng}\ead{zheng@thphys.uni-heidelberg.de}
\author[second]{Ziad Sakr}\ead{sakr@thphys.uni-heidelberg.de}
\author[third]{Luca Amendola}\ead{l.amendola@thphys.uni-heidelberg.de}
\affiliation[first,second,third]{organization={Institut f\"{u}r Theoretische Physik, Universit\"{a}t Heidelberg},%Department and Organization
            addressline={Philosophenweg 16}, 
            postcode={69120}, 
            city={Heidelberg},
            country={Germany}}

\begin{abstract}
We show how one can test the cosmological Poisson equation by requiring only the validity of three main assumptions: the energy-momentum conservation equations of matter, the equivalence principle, and the cosmological principle. We first point out that one can only measure the combination ${\mathcal M}\equiv \Omega_m^{(0)}\mu$, where $\mu$ quantifies the deviation of the Poisson equation from the standard one and $\Omega_m^{(0)}$ is the fraction of  matter density at present. Then we employ a recent model-independent forecast for the growth rate $f(z)$ and the expansion rate $E(z)$ to obtain constraints on ${\mathcal M}$ for a survey that approximates a combination of the Dark
Energy Spectroscopic Instrument (DESI) and Euclid. We conclude that a constant ${\mathcal M}$ can be measured with a relative error $\sigma_{\mathcal{M}}=4.5\%$, while if ${\mathcal M}$ is arbitrarily varying in redshift, it can be measured only to within $13.4\%$ (1 $\sigma$ c.l.) at redshift $z=0.9$, and 15-22\% up to $z=1.5$. We also project our constraints on some parametrizations of ${\mathcal M}$ proposed in literature, while still maintaining model-independence for the background expansion, the power spectrum shape, and the non-linear corrections. Generally speaking, as expected, we find much weaker model-independent constraints than  found so far for such models. This means that the cosmological Poisson equation  remains quite open to various alternative gravity and dark energy models.
\end{abstract}




\end{frontmatter}

%\tableofcontents



\section{Introduction}
\label{sec:intro}

The Poisson equation is one of the fundamental theoretical construct on which most of cosmology and astrophysics is based upon. Although its validity has been confirmed to strong precision in the solar system and, indirectly, on small astrophysical scales, no model-independent test of it has been so far  possible on cosmological scales. In this paper we present a way to achieve this.

When cosmological data are analysed to extract cosmological information, for instance about the cosmic expansion, the perturbation growth rate, or deviations from standard gravity, one usually assumes a set of specific models, for instance  extensions or modifications of the $\Lambda$CDM standard cosmological model. Unavoidably, therefore, the conclusions one derives are, to some extent, valid only for that particular set, and should be obtained anew when a different theoretical set is chosen. One way to alleviate, but not overcome, this dependence is to encapsulate extensions to $\Lambda$CDM in  more and more general parameterisations for, e.g. the equation of state (EoS) evolution for dark energy \cite{Caldwell:1997ii}, the growth factor as function of the growth index $\gamma$ \cite{2005PhRvD..72d3529L} or the evolution of the Newtonian and Weyl potential as a function of  the matter density contrast \cite{Bertschinger:2008zb,Amendola:2012ky}. However these parameterisations as well need to assume a model dependent evolution with time \cite{Planck:2018vyg,DES:2022ccp,Sakr:2023bms}.  
In recent work (\cite{Amendola:2019laa,Amendola:2022vte,Amendola:2023awr}), a way to 
free the data analysis from  this built-in model dependence has been investigated and shown to be able to produce relatively strong constraints of various quantities, including the expansion rate, the distance,  the perturbation growth rate, and the ratio of the two linear gravitational metric potentials, the so-called anisotropic stress $\eta$.
The latter can  be constructed by a suitable combination of quantities extracted from the galaxy clustering, the shear lensing, and the supernovae Ia Hubble diagram, without assumptions concerning the power spectrum shape or evolution, or the bias function.

There is another parameter that is often employed to express the deviation from standard gravity, sometimes denoted as $G_{\rm eff}$ or $\mu=G_{\rm eff}/G_{\rm N}$ (where $G_{\rm eff}=G_{\rm N}$ denotes a standard Poisson equation in Planck units). This parameter quantifies how much the Poisson equation differs from the standard case, and is the focus of this paper. If dark energy does not cluster on the observed scales -- as expected in most models of dark energy (see \cite{{Kunz:2007nn}} for a discussion on this assumption) -- then $\mu$ is indeed a measure of deviation from Einstein's gravity. In a scalar-tensor theory, for instance, $\mu$  can be expressed in terms of an extra interaction between the scalar field and matter \cite{faraoni2004cosmology}. 
 Several papers  tried to constrain $\mu$ with various degrees of  model-independence. In \cite{Raveri:2021dbu} the authors used principal component analysis methods to constrain $\mu$ and the cosmological stress tensor using multiple cosmological probes. However, they also used $f\sigma_8(z)$ measurements obtained from galaxy correlations in a model dependent way assuming $\Lambda$CDM model for the geometrical distances, a fixed shape for the power spectrum entering the $\sigma_8$ and different parameterisations to model the bias for each measurements.  
 
 In \cite{Zhao:2015wqa}, the authors followed similar methods to forecast bounds on $\mu$ and on the dark energy EoS parameter $w$ from future SKA galaxy clustering data. However, they assumed a model for the galaxy bias that is a linear function of the scale factor. Recently, Refs. \cite{Mu:2023zct} and \cite{Ruiz-Zapatero:2022xbv} constrained  $\mu$ with current data for the growth $f\sigma_8(z)$ from redshift space distortions (RSD) and $H(z)$ from cosmic chronometers. These papers  used measurements of the   growth  $f\sigma_8(z)$ obtained assuming a model-dependent shape for the power spectrum; moreover, Ref. \cite{Mu:2023zct}  included a constraint on the matter density that is also model dependent  (we will comment more on this below). Finally, Ref. \cite{Sakr:2023bms} relaxed $\sigma_{8}^{(0)}$ from its model dependency in an attempt to constrain the growth index $\gamma$, a parameter that could be  remapped into $\mu$  (see \cite{Pogosian:2010tj,Sakr:2021ylx}), but assumed  $\Lambda$CDM model for the background evolution. 
 In other works, $\mu$ is derived from specific models \cite{Pettorino_2012,Gomez-Valent:2020mqn,Euclid:2023rjj,Cataneo:2021xlx} or is parameterized in a phenomenological way \cite{Planck:2018vyg,DES:2018ufa,Ferte:2017bpf,DES:2022ccp,Sakr:2021ylx,Casas:2022vik}.
 This brief review of some of the previous work shows that so far all tests of the Poisson equation have relied in significant measure on assumptions on the power spectrum shape, on the background evolution, on the growth factor, or on the bias parametrization.  In many of these works, tight constraints on the Poisson equation are produced, but they are valid only within such restrictive assumptions. In this paper we present a possible way to overcome  these limitations and show that the  cosmological constraints on the Poisson equation are much weaker than so far presumed. 

Before continuing, it is necessary to be more precise about our definition of model-independence. We aim at measuring (or forecasting, for now) quantities in a way that is independent of  the background evolution, of the shape of the power spectrum and of the bias function, and of all the other higher-order bias functions that enter the non-linear perturbation theory (for more details on this, see \cite{Amendola:2023awr}). The three main underlying assumptions that are needed for this approach are the cosmological principle, the energy-momentum conservation, and the equivalence principle. The cosmological principle ensures statistical homogeneity and isotropy and therefore it allows us to employ the Alcock-Paczyński effect to measure the cosmic expansion rate. 
The energy-momentum conservation of matter allows us to use redshift-space distortions to measure the growth rate.
Finally, the principle of equivalence allows us to use the generalized non-linear kernel derived in \cite{DAmico:2021rdb}  and to assume that $\mu$ is a universal quantity, that is, the same for dark matter and baryons. 
We are still  nonetheless making an important simplification, namely that the growth rate $f$ is $k$-independent. 
This is of course not correct in some modified gravity models or in presence of massive neutrinos. In these cases, our results for $f$ should be considered as an average over a range of scales. However, we remark that in most models the $k$-dependence is very small: for instance, Ref. \cite{Kiakotou:2007pz} has shown that $f$ varies by less than 2.5$\%$ in the range $k=0.01-0.1h/$Mpc.
In any case, this limitation can in principle be overcome by employing the same methodology at the price, of course, of a weakening of statistical power. 

The main problem of constraining $\mu$, as we discuss in detail below, is that it is degenerate with the matter fraction $\Omega_m(z)$, the expansion rate $E\equiv H(z)/H_0$ where $H_0$ is the Hubble constant at present, and the linear growth rate 
\begin{equation}
    f\equiv\frac{d\log \delta_m}{d\log a}=\frac{\delta_m'}{\delta_m}
\end{equation}
where $\delta_m$ is the linear density contrast of matter, $a(t)$ is the cosmological scale factor, and we use primes to denote derivatives with respect to $\log a$. So in order to measure $\mu$ one needs to measure $\Omega_m(z), E(z)$ and $f(z)$ in a model-independent way.
A method to achieve this for $E$ and $f$ has been recently proposed in \cite{Amendola:2023awr}, in which a Fisher matrix forecast  was obtained  from forthcoming power spectrum and bispectrum data. Since at the moment, as far as we know, there are no proposals to measure $\Omega_m$ in such a model-independent way (see however \cite{Sakr:2023hrl,Dinda:2023xqx} for an investigation of this aspect), we cannot disentangle $\mu$ from it, and therefore   we constrain the combination $\Omega_{m} \mu$ 
in various redshift bins. Similar degeneracy has also been discussed in \cite{Ruiz-Zapatero:2022xbv}. If we add the standard hypothesis that matter is pressureless, and therefore its density $\rho_m$ scales as $a^{-3}$, then we can measure 
\begin{equation}
    \mathcal{M} \equiv \Omega_{m}^{(0)} \mu(k,z)
\end{equation}
where $\Omega_{m}^{(0)}$ is the present value of the matter density fraction. Additionally, given $\mathcal{M} $ at two different redshift bins, we can as well take their ratio and obtain $\mu(z_2)/\mu(z_1)$. If this ratio is different from unity, the standard Poisson equation is violated (of course, $\mu $ could be redshift-independent; in this case, a constant ratio  would be inconclusive). In this sense, and within the limits of our assumptions, we can state, as advertised in the title of this paper, that our method is indeed a model-independent test of the Poisson equation. 

In this paper we present the general idea and produce a forecast for surveys that approximates a combined DESI \cite{Hahn:2022dnf} and Euclid \cite{laureijs2011euclid} survey. In future work we will apply the method to real data.

%%%%%%%%%%%%%%%%%%%%%%%%%%%%%%%%%%%%%%%%%%%%%
%%%%%%%%%%%%%%%%%%%%%%%%%%%%%%%%%%%%%%%%%%%%%

%%%%%%%%%%%%%%%%%%%%%%%%%%%%%%%%%%%%%%%%%%%%%

\section{Model-independent cosmological observables}\label{sec:method}

We consider the line element for a linearly perturbed Friedmann-Lemaître-Robertson-Walker metric in the longitudinal gauge, with the two scalar degrees of freedom characterized by the gravitational potentials $\Psi$ and $\Phi$,
\begin{equation}\label{psimg}
    ds^2=-(1+2\Psi)dt^2+a(t)^2(1+2\Phi)\delta_{ij}dx^idx^j.\,
\end{equation}
We choose units such that $c= G_{\rm N}=1$.
Here we do not consider vector and tensor perturbations. Assuming a spatially flat Universe and a pressureless perfect fluid for the matter content, one can obtain the two Poisson equations for $\Psi,\Phi$  in Fourier space, from the perturbed Einstein equations  \cite{Amendola:2019laa,Pinho:2018unz}. The gravitational potential $\Psi$, assuming it is slowly varying, can be mapped out through the equation of linear matter perturbations,
\begin{align}\label{eq:k^2Psi}
    k^2\Psi&=-4\pi\mu(k,z)a^2\rho_m(z)\delta_m(k,z), \,
\end{align}
where $k$ is the comoving wavenumber, 
 $\rho_m$ and $\delta_m$ the average of the background matter density and the root-mean-square matter density contrast, respectively.  The dimensionless effective gravitational constant $\mu$, depending in principle both on  time and scale, is the function quantifying modified gravity. In standard General Relativity,  at deep subhorizon scales, one has $\mu=1$.  Here we assume universal gravity, i.e. the equivalence principle is preserved. For other consideration see e.g. the work of \cite{Castello:2022uuu,Gomez-Valent:2022bku}. We also assume pressureless, conserved matter, described as a perfect fluid. Then the Euler and continuity equations at deep subhorizon scales ($ k/(aH)\gg$1), in terms of the perturbed variables,  are given by, 
\begin{align}
\theta_m' & =-\left(2+\frac{H'}{H}\right)\theta_m+({\frac{k}{aH}})^{2}\Psi\label{eq:euler}\\
\delta_m' & =-\theta_m
\end{align}
where $\theta_m=ik_{i}v^{i}/aH$ is the velocity divergence of matter. Combining these two equations we obtain the second order growth equation:
\begin{equation}
\delta_{m}''+\delta_{m}'(2+\frac{E'}{E})=-({\frac{k}{aH}})^{2}\Psi\label{eq:growth}
\end{equation}
Substituting $\delta_m'$ with $f\delta_m$, this becomes 
\begin{equation}\label{eq: lin}
  -k^2\Psi=a^2H^2f\delta_m\left[f+\frac{f'}{f}+\left(2+\frac{E^\prime} {E}\right)\right]   .  \,
\end{equation}
Plugging Eq. (\ref{eq:k^2Psi}) into Eq. (\ref{eq: lin}), and replacing $\rho_m$ with $3H^2 \Omega_m /8\pi$, and assuming $\rho_m$ scales as $(1+z)^3$, we are now able to extract 
\begin{equation}\label{eq:M}
\mathcal{M}(k,z)\equiv\Omega_m^{(0)}\mu(k,z)=\frac{2E^2\left[f^2+f^\prime +f\left(2+\frac{E^\prime}{E}\right)\right]}{3(1+z)^3}, \,
\end{equation}
which shows that ${\mathcal M}$ depends only on $f$ and $E$ and their derivatives. From now on we only consider time dependence for ${\mathcal M}$, $f$ and $E$. Once we have model-independent constraints on $f$ and $E$, it is straightforward to put constraints on $\mathcal{M}$. Even though Eq. (\ref{eq:M}) has been obtained under the assumption of a flat Universe, it is in general  a good approximation also for the non-flat case at sub-horizon scales \footnote{For a non-flat Universe Eq. \eqref{eq:k^2Psi} is replaced by (see e.g.  \cite{Lesgourgues:2013bra}),
\begin{equation} \nonumber
     (k^2-3 K)\Psi = -4\pi\mu(k,z)a^2\rho_m(z)\delta_m(k,z), \,
\end{equation}
in which $K=-H_0^2\Omega_{k}^{(0)}$, and $\Omega_{k}^{(0)}=1-\Omega_{\rm tot}^{(0)}$ is the curvature fractional density  today. Combining this modified Poisson equation with Eq. \eqref{eq: lin}, Eq. (\ref{eq:M}) becomes: 
\begin{equation}\nonumber
    \frac{ \Omega_{m}^{(0)}\mu }{1-3 K/k^2}=\,\frac{2\, E^2\left[f^2+f^\prime +f\left(2+\frac{E^\prime}{E}\right)\right]}{3(1+z)^3} \,.
 \end{equation}
 So in non-flat case the quantity one can constrain is
 \begin{equation} \nonumber
     {\mathcal M}\equiv \frac{\Omega_m^{(0)}\mu}{1-3K/k^2}
 \end{equation}
However, since at sub-horizon scales $k\gg H_0$, the term $3K/k^2$ would be negligible also due to the small $\Omega_k^{(0)}$ constrained from observations in standard models.
}. Notice that we will not take $E(z)$ from standard candle/ruler measurements, because they measure distances, and $E(z)$ can be inferred  from distances only in flat space.

We will first estimate the constraints on ${\mathcal M}(z)$ on individual redshift bins for a  survey that approximates the specifics of DESI from $z<0.6$ and of Euclid in the range $z\in (0.6-2)$.  We call this survey the  DE combined survey. The constraints turn out to be relatively weak. Then, since
several parametrizations for $\mu(z)$ have been proposed in the past, we move to put constraints on them. We consider explicitly three such models.
The first, and perhaps the simplest one, arises when assuming a non-minimal constant coupling $\beta$ between dark energy and matter (see e.g. \cite{Amendola:2019laa}). In this case in fact
\begin{equation}
    \mu=1+2\beta^2 \,.
\end{equation}

The second model is by design sensitive to cosmic acceleration \cite{Planck:2018vyg, DES:2018ufa,Ferte:2017bpf}, and we refer to it as "parametrized dark energy" or PDE. Compared to $\Lambda$CDM, it has only one extra parameter $\mu_0$,  quantifying the departure from the standard Poisson equation in which $\mu_0=0$. In this model $\mu$ follows, 
\begin{equation}\label{eq:mupde}
    \mu=1+\frac{\mu_0}{1-\Omega_m^{(0)}+\Omega_m^{(0)}(1+z)^{3}} \,.
\end{equation}

The third model is the Jordan-Brans-Dicke (JBD) theory of gravity \cite{Brans:1961sx}. It belongs to a class  of general scalar-tensor theories but it is particularly simple since it depends only on one parameter, $w_{\rm BD}$. In the GR limit one has $w_{\rm BD}\rightarrow\infty$. From a well-known solution of Brans-Dicke gravity in \cite{1968PThPh..40...49N,1969PThPh..42..544N,1973Ap&SS..22..231G}, $\mu$ follows a power law:
\begin{equation} \label{mu_JBD}
    \mu=(1+z)^{\frac{1}{w_{\rm BD}\,+1}}\,.
\end{equation}

Finally, we also briefly consider 
the normal-branch of Dvali-Gabadadze-Porrati (nDGP) braneworld gravity (\cite{Dvali:2000hr}).
In this model matter is confined to  a four-dimensional brane embedded in five-dimensional Minkowski spacetime. The cross-over scale $r_c = G_5/(2G_{\rm N})$, with $G_5$ being the Newton's constant in five-dimensional spacetime, is the only extra parameter with respect to GR and characterises the behavior of gravity. On scales above $r_c$  gravity is five-dimensional. In the GR limit one has $r_c\rightarrow\infty$. It is useful to define the cross-over energy density fraction,
\begin{equation}
\Omega_{\rm rc} \equiv 1/4r_c^2H_0^2 \,,
\end{equation}
In this model, the Poisson parameter $\mu$ follows (see e.g. \cite{Euclid:2023rjj}), 
\begin{equation}\label{eq:ndgp}
    \mu(a) = 1+\frac{1}{3\alpha}, \qquad  \alpha \equiv 1+\frac{H}{H_0}\frac{1}{\sqrt\Omega_{\rm rc}}\left( 1+\frac{H'}{3H}\right)  \,.
\end{equation}
However, we find later that due to the strong degeneracy between $\Omega_m^{(0)}$ and $\Omega_{\rm rc}$, the nDGP parameter $\Omega_{\rm rc}$ is essentially unconstrained.

%%%%%%%%%%%%%%%%%%%%%%%%%%%%%%%%%%%%%%%%%%%%%%%%%%%%%%%%%%%%%%%%%%%%%%%%%
%%%%%%%%%%%%%%%%%%%%%%%%%%%%%%%%%%%%%%%%%%%%%%%%%%%%%%%%%%%%%%%%%%%%%%%%%

\section{Methodology}\label{sec:err_est}

In \cite{Amendola:2023awr} a method, called FreePower, has been presented to estimate some cosmological functions regardless of the cosmological model. The method makes use of the galaxy power spectrum with one-loop correction and of the tree-level bispectrum.  We briefly review the main results of Ref. \cite{Amendola:2023awr} since we are going to use them here. Some more information is provided in \ref{sec:fp}.

As shown in \cite{DAmico:2021rdb}, the one-loop power spectrum and the tree-level bispectrum depend on the linear power spectrum shape, on its growth function, and on two time-dependent, but space-independent, bias functions and five so-called bootstrap parameters, also time-dependent. The bootstrap parameters define general kernel functions for the one-loop correction that depend only on the assumption of the equivalence principle, the non-relativistic limit of the GR diffeomorphism (the so-called extended Galilean invariance \cite{DAmico:2021rdb}), and on the conservation of the matter energy-momentum tensor. They do not depend therefore  on a specific cosmological model and can be used therefore to parametrize a vast class of cosmologies, including forms of modified gravity. 
Moreover, we do not rely on a particular shape of the power spectrum (e.g. based on inflationary initial conditions), because in our Fisher matrix approach we take the power spectrum $P(k)$ free to vary in many wavebands. We are also independent of the background cosmology because we leave the expansion rate $E(z)$ and the dimensionless angular-diameter distance $H_0 D_A(z)$ free to vary at every redshift. Finally, as already mentioned, we leave the bias and bootstrap parameters free to vary at every redshift. This approach  frees ourselves from any specific cosmological model, at least within a large class of models.

For what follows, it is in particular important to stress that we leave the growth rate $f(z)$, the dimensionless expansion rate $E(z)$, and the angular diameter distance $D_A(z)$ free.
The Alcock-Paczyński (AP) effect, based upon the assumption of the cosmological principle, allows one then to obtain constraints  on $E(z)$ and  $H_0 D_A(z)$.
It is important to remark that at linear level one can only get the combination $f\sigma_8(z)$ (up to a multiplicative constant), while to constrain $f$ alone the extension to non-linear power spectra and/or bispectra is necessary.
In \cite{Amendola:2023awr} the forecasts for a survey that approximates the Euclid specification \cite{Euclidmission} assuming as fiducial a standard $\Lambda$CDM (the same we assume here) have been obtained. 
Here we follow with some changes the specifications in \cite{Matos:2023jkn} that join a DESI-like survey at low redshift to a Euclid-like one at higher redshift, according to Table \ref{tab:spec}.
The DESI-like survey reproduces the specifications for the DESI Bright Galaxy Survey for $z\le 0.6$ based on \cite{Hahn:2022dnf}, and the DESI Emission Line Galaxies (ELG) survey for $0.6\le z \le 0.9$ based on \cite{DESI:2016fyo}, while for $0.9\le z \le 2.0$ we assume
an Euclid-like survey based on \cite{Amendola:2023awr} and \cite{Yankelevich:2018uaz}. We call this the DE combined survey. 

The fiducials for all parameters are based on $\Lambda$CDM and we refer to \cite{Amendola:2023awr}  and \cite{Matos:2023jkn} for more detailed information.
Since here we need only the constraints on $E$ and $f$, we selected from the resulting full  parameter covariance matrix  only the corresponding entries for $n_b=10$ redshift bins, with size $\Delta z=0.2$ and central redshifts $\bar{z}=\{$0.1, 0.3, 0.5, 0.7, 0.9, 1.1, 1.3, 1.5, 1.7, 1.9$\}$. We adopt the "aggressive" case  \cite{Amendola:2023awr}, which implies $k_{\rm max}=0.25 h/$Mpc for the one-loop power spectrum and $k_{\rm max}=0.1 h/$Mpc for the bispectrum.

\begin{table*}
\centering
\begin{tabular}{ccc}
$ \bar z$ & $V$[Gpc$/h]^3 $ & $  n_{\rm g}\times10^3 $[$h$/Mpc]$^3 $ \\
\hline
 0.1 & 0.26 & 143 \\
 0.3 & 1.53 & 15.9 \\
 0.5 & 3.33 & 1.34 \\
 %0.5 & 3.33 & 0.485 \\
 %0.7 & 5.5 & 1.98 \\
 0.7 & 5.5 & 1.07 \\
 0.9 & 7.2 & 1.54 \\
 1.1 & 8.6 & 0.892 \\
 1.3 & 9.7 & 0.522 \\
 1.5 & 10.4 & 0.274 \\
 1.7 & 11. & 0.152 \\
 1.9 & 11.3 & 0.0894 \\

\end{tabular}
\caption{\label{tab:spec}Bins' central redshift, survey volume, and galaxy density at each bin of the DE combined survey.}
\end{table*}

Taking advantage of the Fisher matrix containing $f_i$ and $E_i$ with $i=0,1,\cdots,9$ corresponding to the redshift bin value in the array defined above,  our goal is to project this $(2\times10)^2=20^2$  matrix onto $\mathcal{M}$ to obtain its constraints. From Eq. \eqref{eq:M}, $\mathcal{M}$ depends on the derivatives $f^\prime$ and $E^\prime$, which can be numerically approximated as $f_i^\prime=(f_{i+1}-f_{i-1})/\Delta N_i$ and $E_i^\prime=(E_{i+1}-E_{i-1})/\Delta N_i$ with $\Delta N_i=\ln[(1+z_{i-1})/(1+z_{i+1})]$, and therefore, the discretized values $\mathcal{M}_i$ are only defined in $n_b-2$ bins. 
Accordingly, we can now discretize \eqref{eq:M}, obtaining $\mathcal{M}$ at each bin,
\begin{equation}\label{eq:n_mu}
\mathcal{M}_i =\frac{2E_i^2\left[f_i^2+\frac{f_{i+1}-f_{i-1}}{\Delta N_i} +f_i\left(2+\frac{E_{i+1}-E_{i-1}}{\Delta N_iE_i}\right)\right]}{3(1+z_i)^3}\,,
\end{equation}
where $i= 1,2,\cdots, n_b-2$. We assume now that $\mathcal{M}_i$ is distributed as a Gaussian variable. The expansion of $\mathcal{M}_i$ around its fiducial  reads 
\begin{equation}
\label{eq:difM}
\mathcal{M}_i= \mathcal{M}_{i(\mathcal{F})}+\sum_{j=1}^{6}\frac{\partial \mathcal{M}_i}{\partial X_j^{(i)}}\biggr|_{\vec{X}^{(i)}_{(\mathcal{F})}}\Delta X_j^{(i)} \,,
\end{equation}
in which for brevity we define  $\vec{X}^{(i)}=\{f_{i-1}, f_i, f_{i+1}, E_{i-1}, E_i, E_{i+1}\}$, and we use the subscript $(\mathcal{F})$ to denote the values at the fiducial. Finally, the elements of the covariance matrix are given by (see Appendix A),
\begin{align} \label{eq:M_sigma}
\begin{split}
\sigma^2_{\mathcal{M}_i\mathcal{M}_j}=&\langle(\mathcal{M}_i-\mathcal{M}_{i(\mathcal{F})})(\mathcal{M}_j-\mathcal{M}_{j(\mathcal{F})})\rangle \\ =&\langle(\sum_{p=1}^{6}\frac{\partial \mathcal{M}_i}{\partial X_p^{(i)}}\biggr|_{\vec{X}^{(i)}_{(\mathcal{F})}}\Delta X_p^{(i)})(\sum_{q=1}^{6}\frac{\partial \mathcal{M}_j}{\partial X_q^{(j)}}\biggr|_{\vec{X}^{(j)}_{(\mathcal{F})}}\Delta X_q^{(j)})\rangle\\
=& \sum_{p=1}^{6}\sum_{q=1}^{6}\frac{\partial \mathcal{M}_i}{\partial X_p^{(i)}}\biggr|_{\vec{X}^{(i)}_{(\mathcal{F})}}\frac{\partial \mathcal{M}_j}{\partial X_q^{(j)}}\biggr|_{\vec{X}^{(j)}_{(\mathcal{F})}}\sigma^2_{X_p^{(i)}X_q^{(j)}}\,. 
\end{split}   
\end{align}

%%%%%%%%%%%%%%%%%%%%%%%%%%%%%%%%%%%%%%%%%%%%%
%%%%%%%%%%%%%%%%%%%%%%%%%%%%%%%%%%%%%%%%%%%%%
%%%%%%%%%%%%%%%%%%%%%%%%%%%%%%%%%%%%%%%%%%%%%
We choose $\Lambda$CDM as our fiducial model, with $\Omega_m^{(0)}=0.32$. In our redshift range,  $f_i$ and $E_i$ for the fiducial cosmology can be approximated with
\begin{align}
f_i(z_i)&\approx\Omega_m(z_i)^{0.545}=\left[\frac{\Omega_m^{(0)}(1+z_i)^3}{\Omega_m^{(0)}(1+z_i)^3+1-\Omega_m^{(0)}}\right]^{0.545}, \, 
\\
E_i(z_i)&= \sqrt{\Omega_m^{(0)}(1+z_i)^{3}+1-\Omega_m^{(0)}}. \,
\end{align}

Since the relation between ${\mathcal M}_i$ and $f,E$ is  non-linear, we also checked all the results numerically. That is, we generated 5,000 samples of $f,E$ distributed as multivariate Gaussian variables according to their covariance matrix (inverse of the Fisher matrix) and their mean (the $\Lambda$CDM fiducial values), and then obtained the variance of ${\mathcal M}_i$ from Eq. \ref{eq:n_mu}. The numerical results confirm that the Gaussian approximation for ${\mathcal M}_i$ and the analytical calculation of the next section are very accurate.

%%%%%%%%%%%%%%%%%%%%%%%%%%%%%%%%%%%%%%%%%%%%%
%%%%%%%%%%%%%%%%%%%%%%%%%%%%%%%%%%%%%%%%%%%%%

%%%%%%%%%%%%%%%%%%%%%%%%%%%%%%%%%%%%%%%%%%%%%
%%%%%%%%%%%%%%%%%%%%%%%%%%%%%%%%%%%%%%%%%%%%%
%%%%%%%%%%%%%%%%%%%%%%%%%%%%%%%%%%%%%%%%%%%%%

\section{Results and discussion}\label{sec:results}

\subsection{Model-independent forecasting on ${\mathcal{M}}(z)$}
\begin{figure*}
\begin{center}
\includegraphics[width=4.6in,height=3.5in]{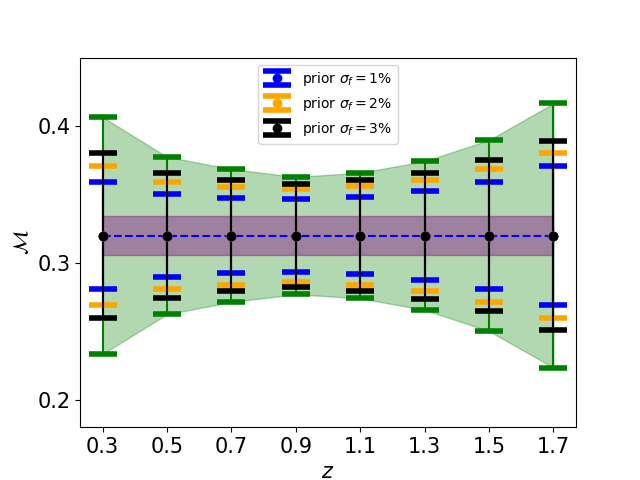}
\caption{ Reconstructed $\mathcal{M}=\Omega_m^{(0)}\mu$ with the 1$\sigma$ estimated error as a function of redshift, with the blue dashed line the fiducial value of $\mathcal{M}=0.32$ at each redshift bin, and the purple area assuming a constant $\mathcal{M}$ along all the bins. Error bars indicate constraints on $\mathcal{M}$ obtained when considering various priors on the growth rate. See the discussion in Sec. \ref{sec:results}.}\label{fig:fig1}
\end{center}
\end{figure*}

\begin{figure*}

\includegraphics[width=6.75in,height=4.5in]{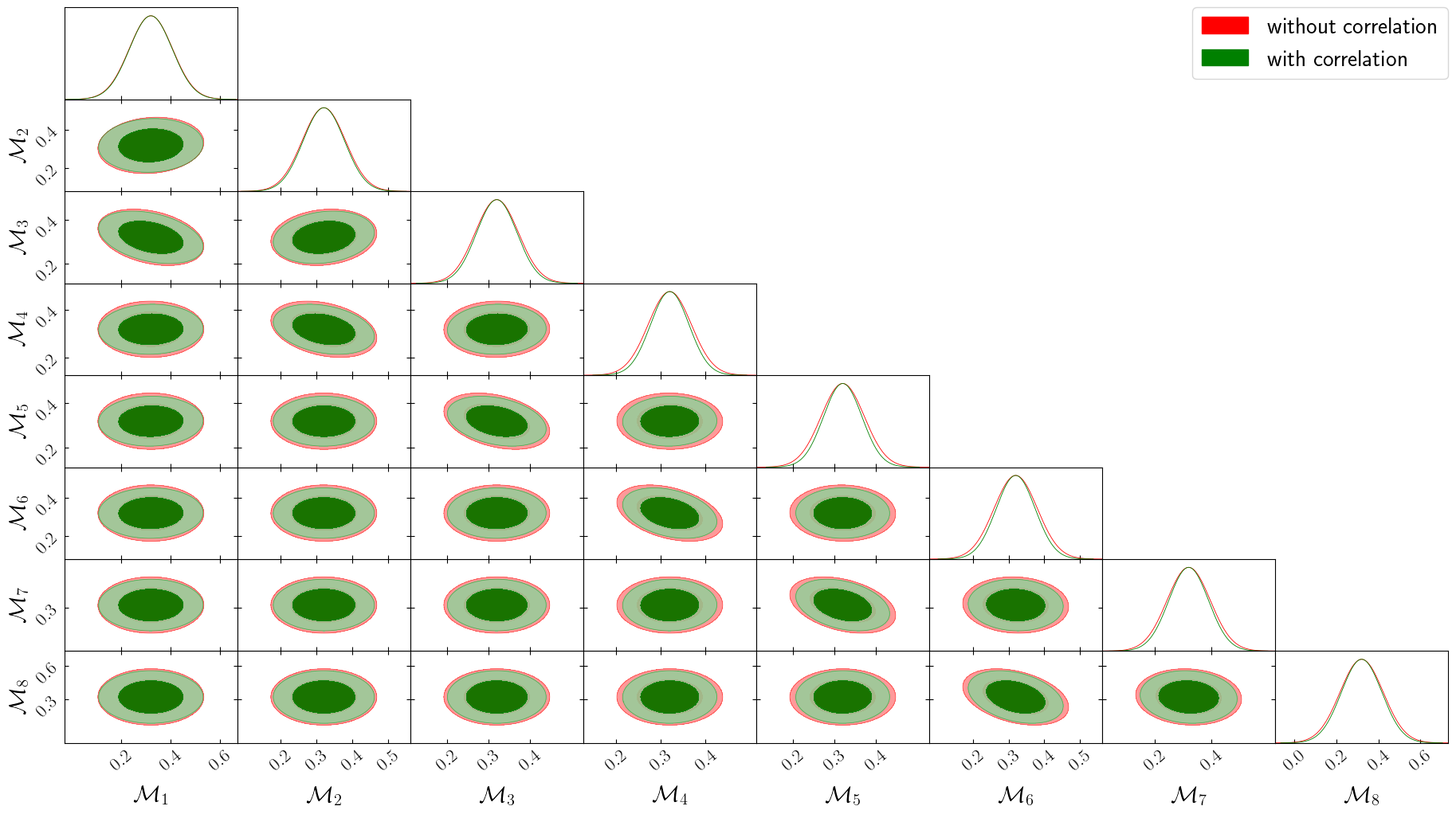}
\caption{Comparison of the one-dimensional posterior distributions of $\mathcal{M}_i$, and the corresponding contour plots at 1$\sigma$ and 2$\sigma$ confidence levels. We observe a negative correlation for each pair of adjacent $\mathcal{M}$'s.}
\label{fig:2D_no_cov}
\end{figure*}

Following the method detailed above, we present the forecasted 1$\sigma$ errors for $\mathcal{M}$  (green shades and lines) in Fig. \ref{fig:fig1} with details on the uncertainty values in each bin for $\mathcal{M}$, $E$ and $f$ in Table \ref{tab:result}. 
%We observe that the error for $\mathcal{M}$ increases with the redshift as expected,
%so that $\mathcal{M}$ is most tightly constrained in the bin $\bar{z}=0.9$,  with $\sigma_\mathcal{M}\approx19\%$.  
We observe that the error for $\mathcal{M}$ increases on both sides of the $z$ axis, as a consequence of the same behavior for the errors on $f$ and $E$ of Table \ref{tab:result}.  To understand this, one has to consider that the Fisher matrix constraints on $f$ and $E$ at linear order are proportional to $\sigma\sim [(n_g P+1)/Vn_g P]^{-1/2}$, where $P$ is the power spectrum. The density $n_g$ is  small at high redshifts, so $n_gP\ll 1$, and the product $Vn_g P$ is smaller than at around $z=1$. Therefore $\sigma$ gets larger. At lower redshifts, instead, $n_g P\gg 1$, so $\sigma\sim V^{-1/2}$, and since the volumes are smaller, $\sigma$ increases again. So we can expect the best constraints somewhere at  intermediate distances.
As we see in Table~\ref{tab:result}, although the errors on $E$ and $f$ are at the percent level, the ones on $\mathcal{M}$ are one order of magnitude higher, reaching more than 30\% for the farthest bins. We include in our analysis the correlations among the various $f$'s,$E$'s, and between the $f$'s and the $E$'s,  both within the same bins and among different bins. It is interesting to highlight that these correlations actually help reducing the errors for $\mathcal{M}$ by more than six percent. In fact, performing the same analysis but removing the correlations, we find relative errors for $\mathcal{M}$ equal to $15.0\%$ and $32.2\%$ in the most tightly constrained and the last bin, respectively. In Fig. \ref{fig:2D_no_cov} we use \texttt{GetDist} \cite{Lewis:2019xzd} to compare posterior distributions for these two cases. Also, as expected from Eq.\ref{eq:n_mu}, we find (negative) correlations between $\mathcal{M}_i$ and $\mathcal{M}_{i-2}$.

In Fig. \ref{fig:fig1} we also report our results with priors. Since $E$ is already strongly constrained around the 1\% level, we explore how the errors on $\mathcal{M}$ would change if we choose strong priors $\sigma_f^{(P)} $ on $f$. These  could be obtained for example if one manages to improve the higher-order correlators and reach a higher $k_{max}$ in the modeling of large scale structure. We show in the same plot in Fig. \ref{fig:fig1} the new errors for $\sigma_f^{(P)} $ equal to 1, 2 or 3\% in blue, orange or black lines, respectively. We observe that the errors improve proportionally to the prior. In the best case, namely with a prior $\sigma_f=1\%$, our constraints on $\mathcal{M}$ tighten by $\approx 37\%$ and the relative error in the bin $\bar{z}=0.9$ is reduced to $8.38\%$.
%\textcolor{red}{(ZS mention the table here as well not only the figure because the latter does not show the relative errors)}. 
  1

\begin{figure*}
\begin{center}
\begin{tikzpicture}
    \node[inner sep=0em] (pic)  at (0,0){\includegraphics[width=4.6in,height=3.5in]{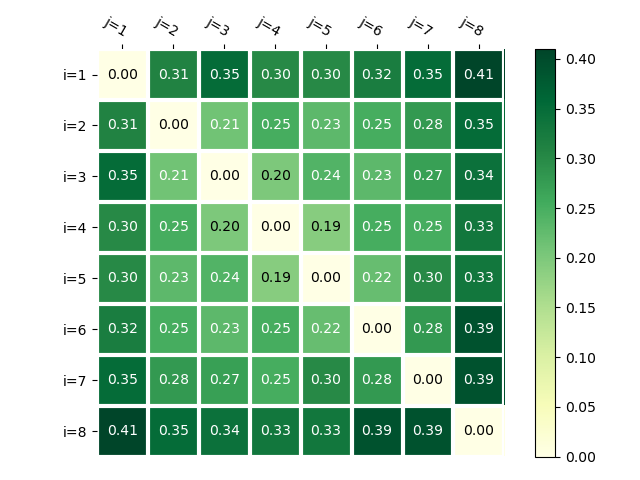}};
    \node[xshift=-1em] at (pic.east) {\large$\sigma_{ij}$};
\end{tikzpicture}
\caption{Visualisation of the errors on $\sigma_{ij}=\sigma(R_{ij})=\sigma(\mu_i/\mu_j) $ at different bins.}\label{fig:ratio.png}
\end{center}
\end{figure*}

\begin{table}
\centering
\begin{tabular}{p{1.0cm} p{0.5cm} p{1.4cm} p{1.4cm} p{1.3cm}}
 $\bar z$ & $i$ &$\sigma_f(\%)$ & $\sigma_E(\%)$ & $\sigma_{\mathcal{M}_i}(\%)$\\ 
 \hline   
 $0.1$ & 0 & 7.44  & 3.03 & - \\ 

 $0.3$ & 1 & 4.44  & 1.74 & 27.04 \\ 

 $0.5$ & 2 & 3.32  & 1.43 & 17.97\\ 

 $0.7$ & 3 & 2.50  & 1.25 & 15.18   \\ 

 $0.9$ & 4 & 2.33  & 1.16 & 13.39\\ 

 $1.1$ & 5 & 2.32  & 1.17 & 14.34\\  

 $1.3$ & 6 & 2.48  & 1.25 & 16.97\\  
 
 $1.5$ & 7 & 3.01  & 1.47 & 21.80\\  
 
 $1.7$ & 8 & 3.85  & 1.82 & 30.16\\  
 
 $1.9$ & 9 & 5.14  & 2.37 & -\\  
\end{tabular}
\caption{   1$\sigma$ relative errors on $f,E$, and $\mathcal{M}$, obtained from model-independent measurements at various $z$ bins, see the description  in Sec. \ref{sec:err_est}.}
\label{tab:result}
\end{table}
%%%%%%%%%%%%%%%%%%%%%%%%%%%%%%%%%%%%%%%%%%%%%%%%%%%%%%%%%%%%%%%%%%%%%%%%%%%%%%%%%%%%%%%%%
We discuss now the ratios $R_{ij}\equiv \mathcal{M}_i/\mathcal{M}_j=\mu_i/\mu_j$. As mentioned in Sec. \ref{sec:intro}, its deviation from unity indicates directly a violation of the standard Poisson equation, without assumptions on $\Omega_m^{(0)}$. Errors on $\sigma_{ij}=\sigma(R_{ij})$ can be easily estimated using error propagation while  also taking the correlations between the nine $\mathcal{M}_i$'s into consideration. 
We visualise their values in Fig. \ref{fig:ratio.png}. In the best case, we find the constraint on $R_{ij}$ around $20\%$. Models with time-varying $\mu$ would be  excluded if their predictions are outside the regions in Table \ref{tab:result} and/or  Fig. \ref{fig:ratio.png}. If one assumes a power law $G_{\rm eff}=\mu\sim a^{\alpha-1}$, then one can relate $R_{ij}$ to the exponent $\alpha$ 
\begin{equation}
    |\frac{\dot G_{\rm eff}}{G_{\rm eff}}|=H
|\frac{G'_{\rm eff}}{G_{\rm eff}}|=(\alpha-1) H \approx H|\frac{\mathcal{M}_{i}-\mathcal{M}_j}{\mathcal{M}_j d_{ij}}|=H|\frac{R_{ij}-1}{d_{ij}}|
\end{equation}
where $d_{ij}=\ln[(1+z_{j})/(1+z_i)]$, and a dot denotes derivative with respect to the cosmic time. Then the relative error on $\alpha$ is $\sigma_\alpha=\sigma_{ij}/d_{ij}$. Therefore  upper limits  on $R_{ij}$ imply  upper limits to the classic test of relativity $|\dot G_{\rm eff}/G_{\rm eff}|$. We will discuss such a power-law $G_{\rm eff}$ in the context of Jordan-Brans-Dicke gravity below. 

%%%%%%%%%%%%%%%%%%%%%%%%%%%%%%%%%%%%%%%%%%%%%%%%%%%%%%%%%%%%%%%
\subsection{Modified gravity models}

We discuss now some parameterizations of modified gravity that have been proposed in the literature and described at the end of Sect.\ref{sec:method}. In all cases, we project our nine constraints on ${\mathcal M}_i$ onto the specific parameters $X_j$ by a transformation Jacobian $J_{ij}\equiv \partial {\mathcal M}_i/\partial X_j$. 
It is important to remark that although we now choose  specific parametrizations for ${\mathcal M}$, we are still general  as far as   the cosmological model is concerned. Therefore we expect our results will be less strong, but more general, with respect to similar analyses that are carried out within specific models. 

\begin{figure*}
\begin{center}
\includegraphics[width=7.0in,height=2.2in]{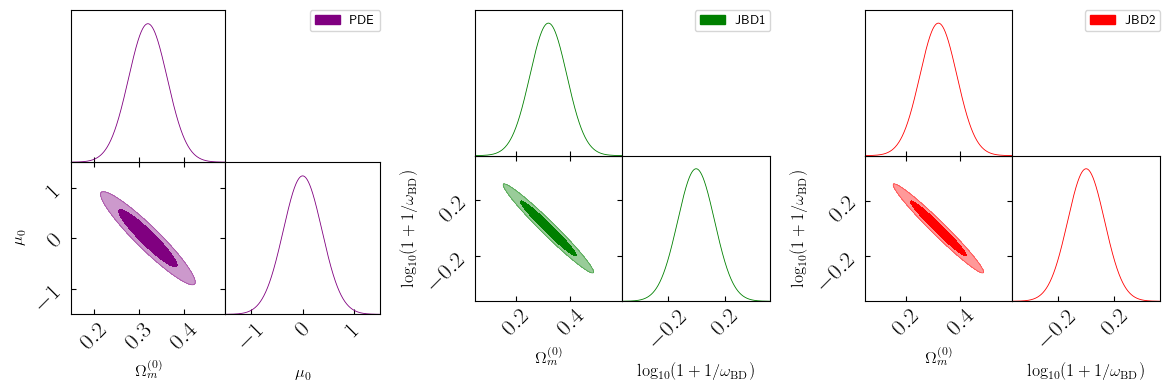}
\caption{One-dimensional posterior distributions of the model parameters varied in the Fisher forecast, and their 2D contour plots at 1$\sigma$ and 2$\sigma$ c.l., for the PDE model and the JBD gravity. Fiducials values are the same as in Table. \ref{tab:table2}.}\label{fig:model}
\end{center}
\end{figure*}

In Table \ref{tab:table2} and Fig. \ref{fig:model} we present our constraints on the aforementioned models. Except for the constant
$\mathcal{M}$ case (in which $\Omega_m^{(0)}$ and $\mu$ are fully degenerate, so we only consider the $\mathcal M$ combination), for each other model, we present our result both varying all the parameters and fixing $\Omega_m^{(0)}$ to 0.32. In all cases, we find strong correlations between $\Omega_m^{(0)}$ and the other modified gravity parameters. Notice that we refrain from adopting a Planck prior, or any other similar prior, because they have been obtained for specific models.

In the first model,  one can assume that $\mathcal{M}$ is constant along all the bins. As already mentioned, such a constant $\mathcal{M}$ is in fact expected in the simplest models with a coupling constant $\beta$ between matter and dark energy. A constant $\mathcal{M}$ will clearly lead to quite stronger constraints than when $\mathcal{M}$ depends freely on redshift. 
The ''new" Fisher matrix $ \textbf{F}^\prime$ (with only one entry for $\mathcal{M}$) is related to the "old" $8\times8$ Fisher matrix for $\mathcal{M}_i$ through the transformation, 
\begin{equation} \label{eq:transformation}
    \textbf{F}^\prime=\textbf{J}^T\,\textbf{F}\, \textbf{J} \,,
\end{equation}
with $\textbf{J}^T=[1,1,1,1,1,1,1,1]$. 
We find a relative error $\sigma_{\mathcal{M}}=4.55\%$. 
%lowered by $\approx 78\%$  compared to model-independent constraint in the bin $\bar{z}=0.9$. 
Since ${\mathcal M}=\Omega_m^{(0)}(1+2\beta^2)$, the relative error for ${\mathcal M} $  corresponds to the absolute error on $2\beta^2$ if we fix $\Omega_m^{(0)}$ to its Planck value. This corresponds to an absolute constraint $\sigma_{\beta^2}\approx 0.0073$ on the coupling $\beta^2$. This constraint is weaker than those from CMB, for instance, Ref. \cite{Gomez-Valent:2020mqn} constrained $\beta=0.0158^{+0.0067}_{-0.0120}$,  which corresponds to $\sigma_{\beta^2}\approx0.0003$, using CMB data alone and without any prior included (for other constraints on $\beta $ see e.g. \cite{Pettorino_2012}). Neglecting the impact of fiducial values,  our error on $\beta^2$ is one order of magnitude larger, but complementary, since it is obtained in a completely different redshift range.

In  the second model, parametrized dark energy, PDE, (see \cite{Planck:2018vyg, DES:2018ufa,Ferte:2017bpf}), the parametrization   is related to $\mathcal{M}$ by (see Eq. \ref{eq:mupde}), 

\begin{table*}
\centering
\begin{tabular}{cc|cc|ccc|ccc}
 & {Constant $\mathcal{M}$} & \multicolumn{2}{c|}{PDE} & \multicolumn{3}{c|}{JBD 1} & \multicolumn{3}{c}{JBD 2}\tabularnewline
& {$\sigma(\mathcal M) $}   & \multicolumn{1}{c}{$\sigma(\Omega_{m}^{(0)})$} & $\;\;\sigma(\mu_{0})$ & \multicolumn{1}{c}{$\sigma(\Omega_{m}^{(0)})$} & \multicolumn{1}{c}{$\;\;\sigma(t)$} & $\;\;w_{\rm BD}$ & \multicolumn{1}{c}{$\sigma(\Omega_{m}^{(0)})$} & \multicolumn{1}{c}{$\sigma(t)$} & $w_{\rm BD}$ \tabularnewline
\hline 
varying all parameters &  0.05 & 0.04 & 0.38 & 0.07 & 0.13 & $> 2.86 $ & 0.07 & 0.13 & $> 2.84$ \tabularnewline

fixing $\Omega_{m}^{(0)}=0.32$ &  - & - & 0.13  & - & 0.03 & $>14.98$ & - & 0.03 & $> 14.68$ \tabularnewline

fiducial values & 0.32  & 0.32 & 0 & 0.32 & 0 & $\infty$ & 0.32 & 5.4 $\times$ $10^{-4}$ & 800 \tabularnewline

\end{tabular}

\caption{First row: forecast 1$\sigma$ marginal absolute errors on all the model parameters. In this table we show results for three models: a constant $\mathcal{M}$, PDE, and the Jordan-Brans-Dicke gravity (with two choices of fiducials). Second row: same, but fixing $\Omega_m^{(0)}=0.32$ and only varying the other parameter. The last row lists the fiducial values used in our forecast.}
\label{tab:table2}
\end{table*}

\begin{equation}
    \mathcal{M}=\Omega_m^{(0)}+\frac{\Omega_m^{(0)}\mu_0}{1-\Omega_m^{(0)}+\Omega_m^{(0)}(1+z)^{3}} \,.
\end{equation}

 We find 1$\sigma$ marginalised constraints $|\mu_0|<0.38$ and $\Omega_m^{(0)}=0.32\pm0.04$ when varying both parameters, with no prior included.  Fixing instead $\Omega_m^{(0)}=0.32$ to the $\Lambda$CDM fiducial, we forecast a constraint $|\mu_0|<0.13$ at 1$\sigma$ c.l. We can compare our result with some model-dependent constraint or forecast in previous works in which a  $\Lambda$CDM background  was assumed. For instance, Ref. \cite{DES:2022ccp} considered  a flat prior on $\Omega_m^{(0)}\in (0.1, 0.9)$ and reported a bound of $\mu_0=0.08^{+0.21}_{-0.19}$ inferred from  measurements from the Dark Energy Survey's first three years of observations, together with external data. Ref. \cite{Sakr:2021ylx} constrains the ratio $\mu_0/\Omega_{\Lambda}^{(0)}=0.11^{+0.38}_{-0.61}$ at 1$\sigma$ c.l. with the Planck data. This results in $\mu_0= 0.076\pm0.26$ neglecting the correlation between $\mu_0 $ and $\Omega_m^{(0)}$. Forecasting on $\mu_0$ has also been done in \cite{Casas:2022vik}, where the authors exploited the  spectroscopic galaxy clustering from the SKA Observatory (SKAO), lying in the redshift range $0 < z < 0.4$. Without  any prior, $\mu_0$ has been constrained as $\mu_0=0.07\pm0.33$ at 1$\sigma$ c.l. Moreover, by combining future surveys from the Vera C. Rubin Observatory (VRO) with SKAO, they found $\mu_0=0.07\pm0.010$. In order to compare with these results, we shifted  our fiducial to $\mu_0=0.07$, and we obtain the constraint $\mu_0=0.07\pm 0.38$. As expected,  model independence considerably opens up the parameter space.

%%%%%%%%%%%%%%%%%%%%%%%%%%%%%%%%%%%%%%%%%%%%%%%
%%%%%%%%%%%%%%%%%%%%%%%%%%%%%%%%%%%%%%%%%%%%%%%

Concerning the third model, Jordan-Brans-Dicke gravity, here we employ two fiducials.  The first one is the $\Lambda$CDM model, namely  $\omega_{\rm BD}=\infty$ (JBD 1); while the other one is  $\omega_{\rm BD}=800$ (JBD 2). The latter is taken from constraints on $\omega_{\rm BD}$ at cosmological scales, typically with a lower  bound of the order of magnitude $10^3$ (\cite{SolaPeracaula:2019zsl, SolaPeracaula:2020vpg, Joudaki:2020shz, Avilez:2013dxa}), which is much smaller compared to solar system constraints, that can reach $\omega_{\rm BD}> 10^5$ (\cite{Bertotti:2003rm,Will:2014kxa}). For this model we perform a  Jacobian projection onto the parameter $t=\log_{10}(1 +1/\omega_{\rm BD})$. This parametrization for $\omega_{\rm BD}$ is a common choice when studying the JBD gravity (see e.g. (\cite{Joudaki:2020shz, Ballardini:2016cvy})). It is related to $\mathcal{M}$ by, 
\begin{equation}
    \mathcal{M}=\Omega_m^{(0)}(1+z)^{\frac{1}{w_{\rm BD}+1}}=\Omega_m^{(0)}(1+z)^{\frac{10^t-1}{10^t}} \,.
\end{equation}
We find weak and almost identical constraints on $\omega_{\rm BD}$ for JBD1 and JBD2, regardless of whether $\Omega_m^{(0)}$ is fixed. We can compare these results with, e.g.,  Ref. \cite{Euclid:2023rjj}, which constrained $\omega_{\rm BD}=800^{+8000}_{-730}$ with the Euclid spectroscopic galaxy survey with no prior included.

Finally, we also expressed the nDGP model parameters in terms of $\mu$ using Eq.~\ref{eq:ndgp}, and projected
 the constraints onto the two parameters $\Omega_m^{(0)}$ and $\rm log_{10}(\Omega_{\rm rc})$. 
Following Ref. \cite{Cataneo:2021xlx}, we choose  $\Omega_{\rm rc}=0.0625$ as fiducial,  and we find $\Omega_{\rm rc}=0.0625_{-0.0523}^{+0.3194}$ fixing $\Omega_{m}^{(0)}$ to 0.32. 
However, if we leave $\Omega_m^{(0)}$ free, but include a wide prior $\Omega_m^{(0)}=0.32\pm 0.1$, the constraints on $\Omega_{\rm rc}$ become totally irrelevant. 

%%%%%%%%%%%%%%%%%%%%%%%%%%%%%%%%%%%%%%%%%%%%%
%%%%%%%%%%%%%%%%%%%%%%%%%%%%%%%%%%%%%%%%%%%%%
%%%%%%%%%%%%%%%%%%%%%%%%%%%%%%%%%%%%%%%%%%%%%

\section{Conclusions}\label{sec:conclusions}
In this work  we have constrained the deviation from the standard Poisson equation in a model-independent way for a future survey that approximates DESI and Euclid in the range $z<2$ and, for the first time, we performed forecasting on the parameter combination $\mathcal{M}=\Omega_m^{(0)}\mu$. Assuming the equivalence principle and the energy-momentum conservation of matter, $\mathcal{M}$ depends only on the growth rate $f$ and on the dimensionless expansion rate $E$, which can be both obtained model-independently from combined power spectrum and bispectrum data. We consider both the cases when $\mathcal{M}$ is a free function of redshift, and when it is parametrized in some forms suggested in the literature.

The main message of this paper is that the Poisson equation can be constrained with near-coming cosmological data only up to 13-30\%, depending on the redshift, so that there is still considerable space for  alternative cosmologies, i.e. models of modified gravity or clustering dark energy.  Only by further assumptions, e.g. that $\mathcal{M}$ is constant, can better bounds be achieved, although always much broader than usually obtained for specific models. This shows that model-independent analyses are a useful tool to discern and quantify the extent to which the constraints on fundamental cosmological quantities depend on specific assumptions.

On the other hand, one could further tighten the constraints on $\mathcal{M}$ by combining measurements from other surveys like SKAO \footnote{https://www.skao.int} and VRO \cite{LSST:2008ijt, LSSTDarkEnergyScience:2012kar, LSSTDarkEnergyScience:2018jkl}, as they involve additional survey volume. Moreover, modeling the non-linear galaxy spectra to higher $k_{\rm max}$ 
will result in more stringent constraints on $E$ and, most importantly, on $f$. In this way, the uncertainty on $\mathcal{M}$ can be significantly lowered. Finally, one could constrain a more general $\mu$ by leaving $f$ free in $k$ and $z$ bins and obtain $\mu(k,z)$.  These developments are left for future work.

\section*{Acknowledgements}
ZS and LA acknowledge support from DFG project  456622116. LA acknowledges  useful discussions with Marco Marinucci, Miguel Quartin, and Massimo Pietroni.

\appendix

\section{ The  covariance matrix}\label{appendA}

The numerical derivatives $\frac{\partial \mathcal{M}_i}{\partial X_j^{(i)}}$ in Eq.~\eqref{eq:difM} are evaluated as follows,
\begin{align}
\begin{split}
\frac{\partial \mathcal{M}_i}{\partial f_j}&=\frac{2E_i^2\left[2f_i + \left(2+\frac{E_{i+1}-E_{i-1}}{\Delta N_iE_i}\right)\right]}{3(1+z_i)^3}\,, \qquad j=i
\\
\frac{\partial \mathcal{M}_i}{\partial f_j} &=\frac{-2E_i^2}{3\Delta N_i(1+z_i)^3}\,, \qquad j=i-1
\\
\frac{\partial \mathcal{M}_i}{\partial f_j} &=\frac{2E_i^2}{3\Delta N_i(1+z_i)^3}\,, \qquad j=i+1
\\
\frac{\partial \mathcal{M}_i}{\partial E_j} &= \frac{4E_i\left[f_i^2+\frac{f_{i+1}-f_{i-1}}{\Delta N_i} +f_i\left(2+\frac{E_{i+1}-E_{i-1}}{\Delta N_iE_i}\right)\right]}{3(1+z_i)^3}-\frac{2f_i(E_{i+1}-E_{i-1})}{3\Delta N_i(1+z_i)^3} \,, \,\,\,\, j=i
\\
\frac{\partial \mathcal{M}_i}{\partial E_j} &=\frac{-2E_if_i}{3\Delta N_i(1+z_i)^3}\,, \qquad j=i-1
\\
\frac{\partial \mathcal{M}_i}{\partial E_j} &=\frac{2E_if_i}{3\Delta N_i(1+z_i)^3}\,, \qquad j=i+1
\end{split}
\end{align} 


\section{The FreePower method}
\label{sec:fp}

The FreePower method has been developed in \cite{Amendola:2023awr}.  We summarize it here for the convenience of the reader, adapting in large measure text already published in \cite{Matos:2023jkn}.

The one-loop power spectrum is defined as
\cite{Ivanov:2019pdj,DAmico:2019fhj}
\begin{align}
    P_{gg}(k,\mu,z) & =S_{{\rm g}}(k,\mu,z)^{2}\left[P^{{\rm lin}}(k,\mu,z)+P^{{\rm 1loop}}(k,\mu,z)+P^{{\rm UV}}(k,\mu,z)\right]+P^{{\rm SN}}(z)\,,\label{Pgg}
\end{align}
Full expressions  are given in Appendix~A of \cite{Amendola:2022vte}  (see also~\cite{Ivanov:2019pdj,DAmico:2019fhj}). The linear matter power spectrum is
\begin{equation}
    P^{{\rm lin}}(k,\mu,z)=Z_1({\mathbf{k}};z)^2 G(z)^2 P_0(k)\,,\label{plin}
\end{equation}
where $P_0(k)$ is the present linear  spectrum  in real space,  and $G(z)$ is the linear growth factor, normalized as $G(0)=1$. The factor $Z_1=b_1+f\mu^2$, where $b_{1}(z)$ is the linear bias parameter and $f(z)\equiv d\log G/d\log a$ is the linear growth rate, represents the redshift space distortion (RSD) correction. 

The non-linear  corrections $ P^{\rm 1loop}$, also called one-loop corrections,
are estimated through integrals of the form
\begin{equation}
   \int P^{\rm lin}({\mathbf{k}_1})(P^{\rm lin}({\mathbf{k}_2})Z_n({\mathbf{k}_1},{\mathbf{k}_2})d{\mathbf{k}_1}d{\mathbf{k}_2}
\end{equation}
where the kernels $Z_n$ depend on the perturbation equations. These kernels
 have been often derived for particular cases, e.g. Einstein-deSitter or $\Lambda$CDM. To enhance the level of model-independence, we adopted however here the approach of \cite{DAmico:2021rdb}, in which they calculate the non-linear correction  with the bootstrap method. This amounts to  imposing some general symmetries (namely, the equivalence principle and the generalized  Galilean invariance), without restriction to specific cosmologies. The bootstrap parameters are represented by five
free $z$-dependent functions, denoted $
    d_\gamma^{(2)}(z)\,,
    a_\gamma^{(2)}(z)\,,\;d_{\gamma a}^{(3)}(z)\,,
\,c_\gamma^{(2)}(z),\,a_{\gamma a}^{(3)}(z)\,.$ Moreover, we include two bias functions, $b_1(z)$ and $b_2(z)$,  a counterterm parameter $c_0(z)$, and shot noise $P^{{\rm SN}}$. The spectrum is also corrected by an  overall smoothing factor $S_{{\rm g}}(k,\mu,z)^{2}$ that takes into account both the Finger-of-God (FoG) effect and the spectroscopic errors~\cite{BOSS:2016psr, Amendola:2022vte}:
\begin{equation}
    S_{{\rm g}}(k,\mu,z) = \exp\left[-\frac{1}{2}(k\mu\sigma_{{\rm z}})^{2}\right] \, \exp\left[-\frac{1}{2}(k\mu\sigma_{f})^{2}\right],\label{eq:FoG}
\end{equation}
where
\begin{equation}
    \sigma_{{\rm z}}=\sigma_{0}(1+z)H(z)^{-1}\,.
\end{equation}
We take  $\sigma_{0}=0.001$ for the spectroscopic errors, and leave the FoG smoothing $\sigma_f$  as a free parameter in each bin.

Beside the one-loop power spectrum, we also include in FreePower the bispectrum at tree-level. We refer to \cite{Amendola:2023awr} for more information. Here is only important to
remark that the same bootstrap kernel functions enter the bispectrum, without new parameters except for two additional shot-noise terms.

The  shape of the power spectrum is
 parametrized by equally spaced wavebands in $\Delta k=0.01 h/$Mpc intervals from 0.01 to 0.25 $h/$Mpc.
    Beside all the bias and bootstrap functions, we include as free parameters the growth rate $f$ and the AP parameters $h\equiv E(z)/E_r$  and $d=D(z)/D_r$ (the subscript $r$ refers to the arbitrary reference values  employed to derive the vector $k,\mu$ from the raw data).

\bibliographystyle{elsarticle-num} 
\bibliography{forecasting_mu,curvature, scaling_bib, references,new-refs}



%% For example:

%% \bibitem[Aladro et al.(2015)]{Aladro15} Aladro, R., Martín, S., Riquelme, D., et al. 2015, \aas, 579, A101

%%\end{thebibliography}

\end{document}